\documentclass[]{aastex63}

\def\lsim{\;\rlap{\lower 2.5pt\hbox{$\sim$}}\raise 1.5pt\hbox{$<$}\;}
\def\gsim{~\rlap{$>$}{\lower 1.0ex\hbox{$\sim$}}}
\newcommand{\chiara}[1]{{\color{black}{#1}}}
\def\lephare{\mbox{\textsc{Le Phare}}}
\def\cigale{\mbox{\textsc{Cigale}}}
\accepted{ApJL}
\shorttitle{Two Einstein crosses from massive post-blue nuggets}
\shortauthors{Napolitano et al.}
\graphicspath{{./}{figures/}}

\begin{document}

\title{Discovery of two Einstein crosses from massive post--blue nugget galaxies at $z>1$ in KiDS\footnote{Based on observations with OmegaCam@VST and MUSE@VLT (Prog. ID: 0105.A-0253).}}

\correspondingauthor{Li, Rui}
\email{lirui228@mail.sysu.edu.cn}

\author[0000-0003-0911-8884]{N.R. Napolitano}
\affiliation{School of Physics and Astronomy, Sun Yat-sen University Zhuhai Campus, Daxue Road 2, 519082—Tangjia, Zhuhai, Guangdong, China}
\affiliation{INAF—Osservatorio Astronomico di Capodimonte, Salita Moiariello 16, I-80131—Napoli, Italy}
\author{R. Li}
\affiliation{School of Physics and Astronomy, Sun Yat-sen University Zhuhai Campus, Daxue Road 2, 519082—Tangjia, Zhuhai, Guangdong, China}


\author{C. Spiniello}
\affiliation{Department of Physics, University of Oxford, Denys Wilkinson Building, Keble Road, Oxford OX1 3RH, UK}
\affiliation{INAF—Osservatorio Astronomico di Capodimonte, Salita Moiariello 16, I-80131—Napoli, Italy}

\author[0000-0001-7958-6531]{C. Tortora}
\affiliation{INAF—Osservatorio Astronomico di Capodimonte, Salita Moiariello 16, I-80131—Napoli, Italy}
\affiliation{INAF—Osservatorio Astronomico di Arcetri, L.go E. Fermi 5, I-50125—Firenze, Italy}


\author{A. Sergeyev}
\affiliation{Institute of Astronomy, V. N. Karazin Kharkiv National University, 35 Sumska Str., Kharkiv, Ukraine }
\affiliation{Institute of Radio Astronomy of the National Academy of Sciences of Ukraine}

\author{G. D'Ago}
\affiliation{Instituto de Astrofísica, Pontificia Universidad Católica de Chile, Av. Vicuña Mackenna 4860, 7820436 Macul, Santiago, Chile}

\author[0000-0002-2338-7709]{X. Guo}
\affiliation{School of Astronomy and Space Science, Nanjing University, Nanjing, Jiangsu 210093, China}
\affiliation{Key Laboratory of Modern Astronomy and Astrophysics (Nanjing University), Ministry of Education, Nanjing 210093, China}

\author{L. Xie}
\affiliation{School of Physics and Astronomy, Sun Yat-sen University Zhuhai Campus, Daxue Road 2, 519082—Tangjia, Zhuhai, Guangdong, China}

\author{M. Radovich}
\affiliation{INAF - Osservatorio Astronomico di Padova, vicolo dell'Osservatorio 5, I-35122 Padova, Italy}

\author{N. Roy}
\affiliation{School of Physics and Astronomy, Sun Yat-sen University Zhuhai Campus, Daxue Road 2, 519082—Tangjia, Zhuhai, Guangdong, China}
\author{L. V. E. Koopmans}
\affiliation{Kapteyn Astronomical Institute, University of Groningen, P.O.Box 800, 9700AV Groningen, the Netherlands}

\author{K. Kuijken}
\affiliation{Leiden Observatory, Leiden University, P.O.Box 9513, 2300RA Leiden, The Netherlands}

\author[0000-0002-3910-5809]{M. Bilicki}
\affiliation{Center for Theoretical Physics, Polish Academy of Sciences, al. Lotników 32/46, 02-668, Warsaw, Poland}

\author{T. Erben}
\affiliation{Argelander-Institut für Astronomie, Auf dem Hügel 71, 53121 Bonn / Germany}

\author{F. Getman}
\affiliation{INAF—Osservatorio Astronomico di Capodimonte, Salita Moiariello 16, I-80131—Napoli, Italy}

\author{C. Heymans}
\affiliation{Institute for Astronomy, University of Edinburgh, Royal Observatory, Blackford Hill, Edinburgh, EH9 3HJ, UK}
\affiliation{Ruhr-University Bochum, Astronomical Institute, German Centre for Cosmological Lensing, Universitätsstr. 150, 44801 Bochum, Germany}

\author{H. Hildebrandt}
\affiliation{Ruhr-University Bochum, Astronomical Institute, German Centre for Cosmological Lensing, Universitätsstr. 150, 44801 Bochum, Germany}

\author{C. Moya}
\affiliation{Instituto de Astrofísica, Pontificia Universidad Católica de Chile, Av. Vicuña Mackenna 4860, 7820436 Macul, Santiago, Chile}

\author{H.Y. Shan}
\affiliation{Shanghai Astronomical Observatory (SHAO), Nandan Road 80, Shanghai 200030, China} \affiliation{University of Chinese Academy of Sciences, Beijing 100049, China}

\author{G. Vernardos}
\affiliation{Institute of Astrophysics, Foundation for Research and Technology$-$ Hellas (FORTH), GR-70013, Heraklion, Greece}

\author{A.H. Wright}
\affiliation{Ruhr-University Bochum, Astronomical Institute, German Centre for Cosmological Lensing, Universitätsstr. 150, 44801 Bochum, Germany}

%




\begin{abstract}
We report the discovery of two Einstein Crosses (ECs) in the footprint of the Kilo-Degree Survey (KiDS): KIDS J232940-340922 and KIDS J122456+005048. Using integral field spectroscopy from MUSE@VLT, we confirm their gravitational-lens nature. In both cases, the four spectra of the source clearly show a prominence of absorption features, hence revealing an evolved stellar population with little star formation.  The lensing model of the two systems, assuming a singular isothermal ellipsoid (SIE) with external shear, shows that: 1) the two crosses, located at redshift $z=0.38$ and 0.24, have Einstein radius $R_{\rm E}=5.2$ kpc and 5.4 kpc, respectively; 2) their projected dark matter fractions inside the half effective radius are 0.60 and 0.56 (Chabrier IMF); 3) the sources are ultra-compact galaxies, $R_{\rm e}\sim0.9$ kpc (at redshift $z_{\rm s}=1.59$) and $R_{\rm e}\sim0.5$ kpc ($z_{\rm s}=1.10$), respectively. {These results are unaffected by the underlying mass density assumption.} Due to size, blue color and absorption-dominated spectra, corroborated by low specific star-formation rates derived from optical-NIR \chiara{spectral energy distribution} fitting, we argue that the two lensed sources in these ECs are blue nuggets migrating toward their quenching phase. 
\end{abstract}



\section{Introduction} \label{sec:intro}
\chiara{Strong} gravitational lensing is a powerful tool to measure the distribution of Dark Matter (DM) in galaxies and study the properties of high-redshift sources. According to General Relativity, matter in the universe acts as a cosmic telescope deforming and magnifying
the light of objects which would be hardly observable otherwise. 
Depending on the size of the source and the alignment of the foreground galaxy (lens or deflector) and the source, strong lensing events show-up as 
arcs or rings (when the source is extended, e.g. a galaxy) or as multiple images (when the source is a compact system, e.g. a quasar). 
Deformed images of background galaxies can be used, in combination with the dynamical and stellar population analysis of the deflector, to determine the lens total mass density profiles (e.g., \citealt{2006ApJ...649..599K}, \citealt{2010ApJ...724..511A}, \citealt{2019MNRAS.489.2049N}), to separate the dark from the luminous matter and to constrain the lens stellar initial-mass-function (IMF) slope (e.g., 
\citealt{2010ApJ...709.1195T}
, \citealt{2011MNRAS.417.3000S}).
Furthermore, doubly (`doublets') or even quadruply\chiara{-lensed} (`quads') QSOs 
are particularly valuable for cosmology because they enable measuring the Hubble constant ($H_0$, \citealt{2013ApJ...766...70S})
via time-delays among the QSO light curves.


However, 
doublets and quads can also be produced by other compact sources, e.g. high--redshift, starforming, ultra--compact galaxies (e.g. \citealt{2012ApJ...761..142M}), which 
are fairly common at high--$z$. Using lensing forecasts from typical ground based surveys \citep{2015ApJ...811...20C} with a depth of the order of 
$r\sim25$, the number of expected quads 
from compact galaxies ($R_{\rm e}<1$ kpc) at redshift $z\lsim4$, with an alignment sufficient to make a cross-like geometry 
(e.g. source misalignment $<$0.1$''$), is of the order of half a dozen every 1000 deg$^2$ (see also \S\ref{sec:conclusions}). Unfortunately, only few of such systems have been observed so far:
besides the \citet{2012ApJ...761..142M} system,
only two Einstein Cross (EC) configurations from Ly-$\alpha$ emitters have been confirmed and fully analysed
(\citealt{2006ApJ...646L..45B}, \citealt{2019ApJ...873L..14B}).  
ECs are interesting {\it per se} as they are the rarest and most spectacular manifestation of quad systems, showing a distinctive symmetric cross pattern around the deflector, 
generated 
when the source and the lens are almost perfectly aligned. Generally, these systems have been found to be produced by distant quasars (e.g. \citealt{1988Natur.334..325M}, \citealt{2018MNRAS.473L.116O}) 
or supernovae (aka Refsdal system, \citealt{2015Sci...347.1123K}). 

Within the Kilo-Degree Survey (KiDS, \citealt{2015A&A...582A..62D}), we have undertaken a systematic search \chiara{for strong} 
gravitational lenses, \chiara{both} 
arcs (\citealt{2017MNRAS.472.1129P}, P+17 hereafter, \citealt{2019MNRAS.484.3879P,2019MNRAS.482..807P}) and multiple images
\citep{2018MNRAS.480.1163S,2019MNRAS.483.3888S}. 
In particular, in the process of improving the overall efficiency of the Convolutional Neural Network (CNN) finders, started with P+17, we have collected, in an area of $\sim$1000deg$^2$, a series of high quality quad candidates \citep[Li+20 hereafter]{2020ApJ...899...30L}, among which we found three clear EC configurations. 

In this paper we report the results of the spectroscopic follow-up of the best two of them.
We argue that these represent a new class of sources of EC configurations, i.e. high$-z$ post-blue nugget systems, and discuss the possibility to systematically search for these objects in current and future ground and space surveys.
For all calculations, we assume a $\Lambda$CDM cosmology with ($\Omega_M$, $\Omega_\Lambda$, $h$)=(0.3, 0.7, 0.7).

\section{Confirmation and Lensing model} \label{sec:confirmation}
The first EC, KIDS J232940-340922 (KIDS-EC1,  hereafter, see Fig. \ref{fig:spectra}), has been found in the Southern KiDS patch. The deflector has a AB magnitude of $r\sim19.8$ and red color, $g-i=1.4$, while the average magnitude of the 4 lensed images of the source is $r\sim22.6$ {and their average color is  $g-i=-0.21$. The second EC, KIDS J122456+005048 (KIDS-EC2, hereafter, Fig. \ref{fig:spectra}), has been found in the Northern KiDS patch. The deflector has total magnitude $r\sim19.7$ and color $g-i=1.9$, while the 4 images have an average magnitude $r\sim22.0$ and average color $g-i=1.0$, 
i.e. bluer than the deflector but redder than KIDS-EC1 source.
The two sources have obtained a high CNN probability and also high human visual score (see Li+20), hence they have been selected for the spectroscopic follow-up.
Table \ref{tab:EC_phot} lists coordinates, relative positions of lens and source images and the optical and near-infrared (NIR) photometry in the $ugriZYJHK_{\rm s}$ bands, for both ECs.} 
To minimize the relative contamination and derive homogeneous photometry for all sources, the 9-band photometry of lens and lensed images are derived by a simultaneous seeing convolved S{\'e}rsic plus 2D Gaussian fit of the objects, respectively, from KiDS-DR4 \citep{2019A&A...625A...2K} and VIKING \citep{2013Msngr.154...32E} calibrated images.

\subsection{MUSE spectroscopy and lensing confirmation}\label{sec:MUSE}
Spectroscopic observations have been collected under ESO Directory Discretionary Time (program ID: 0105.A-0253, PI Napolitano) with MUSE at VLT, Cerro Paranal. Run A has been completed in November 2019 for KIDS J232940-340922 and Run B on February 2020 for KIDS J122456+005048.
Observations have been taken in service mode, in 
wide-field non adaptive optic configuration, 
which \chiara{allows }
a full $1'\times1'$ field of view, in the wavelength range $\lambda=[4750,9300]$\AA. The MUSE grating spectral resolution varies from 1750 to 3750, 
end-to-end, in the same interval.
The total exposure time for both targets is 130 minutes divided in 3 observing blocks (OBs).
Every OB is split in 2$\times1300s$ exposures with a 90deg position angle offset. 
The final seeing of the combined exposures is 0.93$''$ for KiDS-EC1 and 0.84$''$ for KiDS-EC2. 

\begin{figure}
    \centering
    \includegraphics[width=17cm]{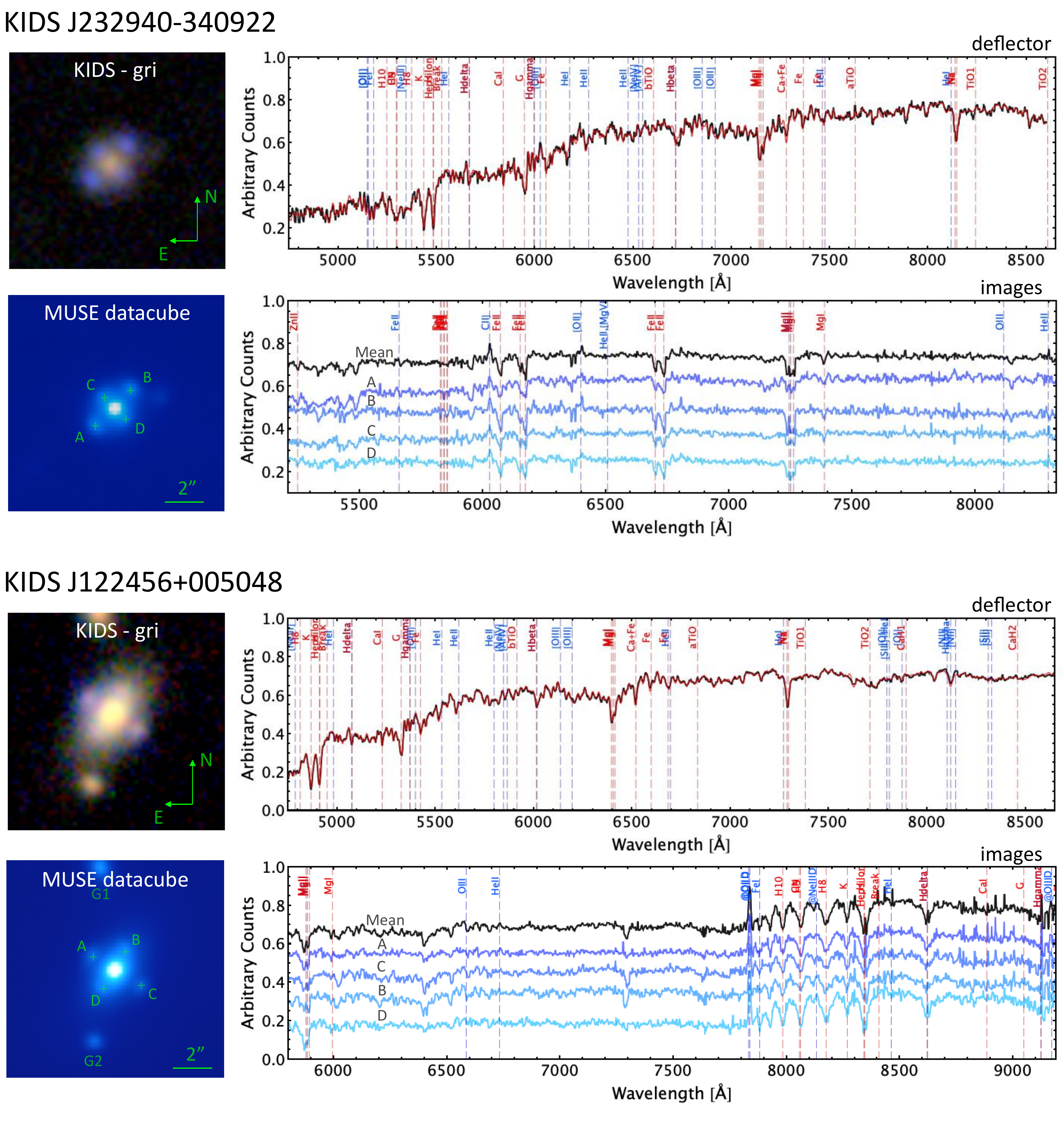}\vspace{3pt}
    \caption{Detection and confirmation of the Einstein crosses: KIDS J232940-34092 (top), KIDS J122456+005048 (bottom). Left column: for each cross we show the KiDS ($gri$) color image (top) and the MUSE white light image (bottom). Right column: for each cross \chiara{we plot} the MUSE spectrum of the deflector, in black, with the best fit model for the velocity dispersion estimate with \textsc{pPXF}, in red, (top) and of the 4 individual images, in blue tones, and the mean spectrum, in black (bottom). Over-plotted on all spectra, are the main absorption lines (in red) and emission lines (in blue), shifted to the estimated redshift of the represented object (see Table \ref{tab:lens_mod}).}
    \label{fig:spectra}
\end{figure}

Reduced data have been provided by ESO as Internal Data Products, using the official MUSE pipeline (v2.8).
For KiDS-EC1 we have also performed our own data-reduction to check consistency with the ESO Phase 3 data products, using the same pipeline\chiara{. We}
have found very consistent spectral quality, in terms of flat fielding, signal-to-noise-ratio (SNR) and sky subtraction. 
\chiara{A zoom-in of the field-of-views (FOVs) of the EC datacubes (i.e. the integrated flux over all wavelengths)} are reported in Fig. \ref{fig:spectra}, and clearly show all sources seen in the KiDS $gri$ color images (also shown in the same figure).
In Fig. \ref{fig:spectra}, we also present the 1D spectra from the four lensed images, {extracted from a single MUSE pixel}, 
and the one of the corresponding deflector, {extracted over an aperture of 3 pixels for KIDS-EC1 and 4 pixels for KIDS-EC2 respectively, corresponding to about half of the effective radius, $R_{\rm e}/2$, see \S\ref{sec:lens_model}}. We use these apertures because they allow us to minimise the
contamination from the blue lensed sources.
For both systems, the lensing nature is confirmed by the presence of identical spectral features in the four different images, all consistent with the same redshift, higher than the one of the lens. For KIDS-EC1, we infer a redshift of $z=1.590\pm0.001$ from Fe and Mg absorption lines and faint [FeII] ($\lambda=5650$\AA), [CII] ($\lambda=6025$\AA) and [OII] ($\lambda=6400$\AA) emissions.
For KIDS-EC2, we calculate instead $z=1.102\pm0.001$, from the spectrum at  $\lambda>7800$\AA, including Balmer absorption lines (H10 at $\lambda\sim 7980$\AA, H9 at $\lambda\sim 8060$\AA, H8 at $\lambda\sim 8180$\AA, H$\epsilon$ at $\lambda\sim 8350$\AA\ and H$\delta$ at $\lambda\sim 8620$\AA), the Calcium K and H doublet (at $\lambda\sim 8360-8450$\AA), although with some sky contamination in H, and a clear [OII] doublet emission line at $\lambda\sim 7850$\AA. In this latter case Fe and Mg absorption lines are also present but look slightly blue-shifted, possibly due to some {gas outflow from the source galaxy (see e.g. \citealt{2014ApJ...794..156R,2020arXiv200503017B})}, which we will investigate in a forthcoming paper.

The two deflectors reveal typical features of early-type galaxies (ETGs), in particular 
a strong rest-frame break at 4000\AA, \chiara{faint Balmer lines in absorption,} and Fe, Mgb and NaD lines, {all} characteristic of an old, metal rich stellar population.
The inferred {deflector} redshifts are $z_{\rm l}=0.381\pm0.001$ for KIDS-EC1, and $z_{\rm l}=0.237\pm0.001$ for KIDS-EC2.
The \chiara{SNR} 
and resolution of the spectra allow us to estimate the velocity dispersion of these two systems using the \textsc{pPXF} software \citep{2017MNRAS.466..798C}, yielding  $\sigma_{R_{\rm e}/2}=192\pm4$ km~s$^{-1}$ for KIDS-EC1 and $\sigma_{R_{\rm e}/2}=248\pm2$ km~s$^{-1}$ for KIDS-EC2\footnote{We estimate that systematic errors from template mismatch and masked regions may amount to $\sim20$ km~s$^{-1}$.} (see also Table \ref{tab:lens_mod}).
The best-fit models are overlaid on the galaxy spectra in the same Fig. \ref{fig:spectra}. 

\subsection{Lensing model, dynamical masses, dark matter fractions}
\label{sec:lens_model} 
\begin{figure}
    \centering
    \includegraphics[width=18cm]{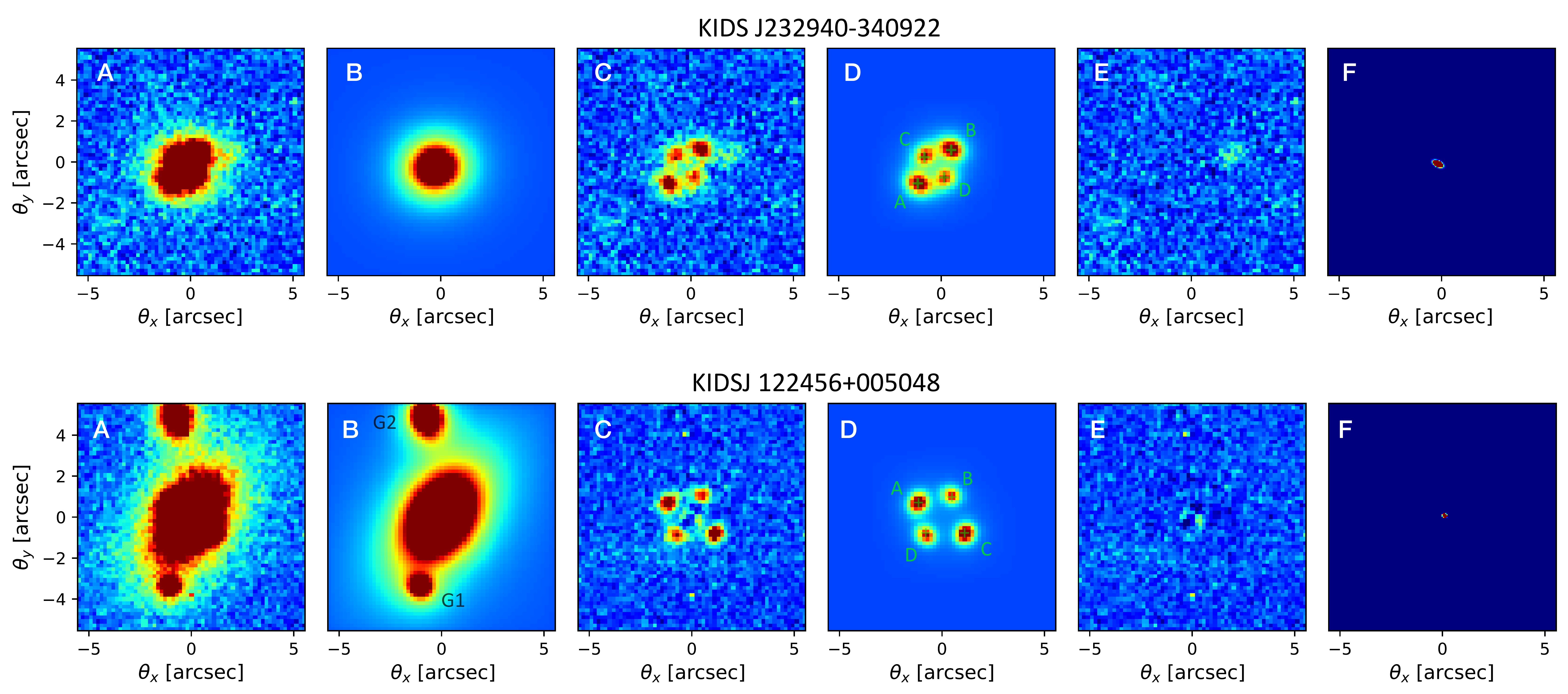}\vspace{3pt}
    \caption{Lensing models for KIDS J232940-34092 (top), KIDS J122456+005048 (bottom). From left to right we show the $r-$band KiDS image used for the model (A), the foreground light-subtracted image (B), the Einstein cross images with the foreground light subtracted (C), the reconstructed Einstein crosses, the residual image (E=A-B-D), and the reconstructed background source (F). We use a singular isothermal ellipsoid to model the deflector mass model and a S{\'e}rsic model for foreground and source light.  For KIDS J122456+005048, to model the foreground light, we also \chiara{account} 
    for the light of the two nearby galaxies marked as G1 and G2 (see panel B). They are also visible in the KiDS and MUSE images (see Fig. \ref{fig:spectra}) and found to have a similar redshift as the lens.}
    \label{fig:lens_mod}
\end{figure}

The two ECs are modelled using the \textit{lfit$\_$gui} code
\citep{2016ApJ...833..264S}.
We use KiDS $r-$band \chiara{images, with }
a pixel scale of $0.2''$ and seeing $0.8''$ for KIDS-EC1 and $0.7''$ for KIDS-EC2. 
The effect of the seeing is taken into account by
convolving the lensing models with a point spread function (PSF) generated by nearby stars \citep[see e.g.,][]{2018MNRAS.480.1057R}. 
The \textit{lfit$\_$gui} code simultaneously models the \chiara{deflector} 
light, the lensed image positions and \chiara{their} magnification, and the best position and light distribution of the source galaxy. 
We assume a single S{\'e}rsic profile \citep{1963BAAA....6...41S} for the two deflectors, although for KIDS-EC2 we need to account also for the presence of two nearby galaxies (G1 and G2, respectively, see Fig. \ref{fig:spectra}), as well as for the two sources. 
The deflector total mass distributions are modelled with a singular isothermal ellipsoid \citep[SIE,][]{1994A&A...284..285K} profile with a projected two-dimensional surface mass density profile described by:
\begin{equation}
\Sigma(x, y)=\frac{1}{2}\Sigma_{\rm c}\, \sqrt{q}\, \theta_{\rm E}\, (x^2+q^2y^2)^{-1/2},
\end{equation}
where $\theta_{E}$ is the lensing strength, equivalent to the Einstein radius, $q$ is the minor-to-major axis ratio of the isodensity contours.  $\Sigma_{\rm c}=c^2/(4\pi G) (D_{\rm s}/D_{\rm d} D_{\rm ds})$ is the critical density, where $D_{\rm s}$ and $D_{\rm d}$  are the angular diameter distances from the observer of the lens and the source, respectively, and $D_{\rm ds}$, the distance between deflector and source. The assumption of a SIE model is motivated by evidence pointing toward 
a logarithmic mass-density slope close to $-2$ 
for the total mass
density around the Einstein radius (e.g. \citealt{2006ApJ...649..599K}), {however the impact of this assumption is discussed in Appendix, while a 
full modeling with more general density profiles will be presented in future detailed analyses}.
We also include external shear, $\gamma_{\rm ext}$, which approximates the influence of the surrounding environment on the lensing potential. 
After having initialized the model, with some test runs,
the final best-fit of the two ECs are obtained
via $\chi^2$ minimization
using the Levenberg–Marquardt algorithm \citep{1978LNM...630..105M}.
These are shown in Fig. \ref{fig:lens_mod}, and the corresponding parameters are reported in Table \ref{tab:lens_mod}. 
From the Table, we can {draw some first general results: }
1) the two ECs show similar Einstein radii, $R_{\rm E}$, {which are both} 
very close to the lens effective radii, $R_{\rm e}$ (i.e. $R_{\rm E}/R_{\rm e}\sim 0.9$); 2) the 
stellar velocity dispersion 
measured from the spectrum of  KIDS-EC1 is smaller \chiara{than that }
inferred by the lens model, suggesting that the actual slope of the total density profile might deviate from $-2$ (see e.g. \citealt{2010ApJ...724..511A} {and the discussion in the Appendix}), while \chiara{the two values are }
consistent within 2$\sigma$ for KIDS-EC2; 3) for both lenses, the total mass may be rounder than the starlight distribution (i.e. $b/a$=0.91 vs. 0.89 for KIDS-EC1 and $b/a$=0.68 vs. 0.59 for KIDS-EC2), but consistent within the errors, {as earlier found in other studies (e.g. \citealt{2020arXiv200811724S} and reference therein)}; 4) for KIDS-EC1 we measure a quite strong external shear ($\gamma_{\rm ext}\sim 0.25$), compatible with a group/cluster potential which is confirmed by the presence of 12 more galaxies at a similar redshift as the lens within $30''$ distance in the MUSE FOV (Napolitano et al. in prep.); 5) both sources are aligned with the lens center of mass within 0.1$''$, while the stellar and mass centers are consistent within the errors.   
{In Appedix we discuss the impact of more general model density assumptions and show that these do not impact the main conclusions of this study.}
From the lens model parameters, we infer a projected mass within $R_{\rm E}$ 
of $\log M(R_{\rm E})/M_\sun=11.28\pm0.02$ for KIDS-EC1 and $\log M(R_{\rm E})/M_\sun=11.42\pm0.01$ for KIDS-EC2 (see also Table \ref{tab:lens_mod}). Due to the underlying assumption of a SIE mass distribution, we can easily derive the mass inside $R_{\rm e}/2$ (i.e., $M(R_{\rm e}/2)$ in Table \ref{tab:lens_mod})
to compare with the dynamical mass by the velocity dispersion measurements inside the same radius, 
derived above.
We remark here that the adoption of
$R_{\rm e}/2$ as reference radius is consistent with previous strong lensing studies at the same scale (e.g. \citealt{2010ApJ...724..511A}).

For the dynamical masses, we use the projected solution of the Jeans Equation inside a circular aperture (e.g. \citealt{2009MNRAS.396.1132T}), in order to take correctly into account the light profile of the lenses (i.e. $n-$index)\footnote{This is the most accurate way to determine the mass inside an aperture and avoid assumptions about the virial estimates for a non-de Vaucouleurs profile ($n\neq4$), see e.g. 
\citealt{2006MNRAS.366.1126C}}.
These projected masses, $M_{\rm J}$ in 
Table~\ref{tab:lens_mod}, are fully consistent with the equivalent lensing-derived masses, hence confirming the self-consistency of our mass estimates of the two systems. 

We finally estimate the total stellar mass, $M_*,$ of the two lens systems via 
SED fitting of the 9-band photometry, given in Table~\ref{tab:EC_phot}. 
We use the public SED fitting--Code Investigating GALaxy Emission (\cigale\ v2018.0, \citealt{2019A&A...622A.103B}. The $M_*,$ are used, in combination with the lensing masses, to derive the DM fraction inside $Re/2$, $f_{\rm DM}=1-M_*(R_{\rm e}/2)/M(R_{\rm e}/2)$. 
We assume solar metallicity, while all other parameters, such as the e-folding time, age of the main stellar population and internal extinction, E(B-V), are free to vary. {For the star formation rate (SFR), we adopt a delayed star formation history, which allows us to efficiently model both typical early-type and late-type galaxies (see \citealt{2019A&A...622A.103B}).} To double check the results we also use
another independent code, \lephare\ (\citealt{2006A&A...457..841I}), with a similar set-up and found consistent results for all the constrained parameters, within the errors.
The stellar mass inside $R_{\rm e}/2$ is derived assuming a constant stellar mass-to-light ratio
for each lens S{\'e}rsic light profile, whose parameters have been inferred from the lensing model in Table~\ref{tab:lens_mod}. The final estimates of $M_*(R_{\rm e}/2)$ are reported in the same Table, where we also list the final $f_{\rm DM}=0.60\pm0.07$ for KIDS-EC1 and $f_{\rm DM}=0.56\pm0.04$ for KIDS-EC2, 
and report the difference between the lensing and dynamical DM fractions
($\Delta f_{\rm DM}$). 

The inferred $f_{\rm DM}$ are typical of DM dominated systems and consistent with previous estimates based on lensing (\citealt{2010ApJ...724..511A}, \citealt{2010ApJ...721L...1T}, \citealt{2019A&A...631A..40S}) or dynamics of local (\citealt{2009MNRAS.396.1132T}, 
\citealt{2012Natur.484..485C}, \citealt{2012MNRAS.425..577T}) or higher redshift galaxies (\citealt{2014ApJ...789...92B}, \citealt{2018MNRAS.473..969T}) of similar stellar masses.

\section{Characterization of the sources} 
\label{sec:char_source}
A striking outcome of the lensing model is the degree of compactness of the source galaxies in both ECs, that turned out to have effective radii smaller than 0.1$''$. The lensing model also provides an estimate of the $n-$indexes ($<1$ in both cases) and axis ratios ($\sim 0.9$ and $\sim0.6$ respectively), suggesting that the two sources might be disk dominated systems\footnote{The best-fit $n-$indexes are fairly small if compared to typical disks: however we have checked that fixing $n=1$, the other parameters change within the errors and the reduced $\chi^2$ is worsened, hence demonstrating that $n-$indexes are realistically $\lsim1$ (see e.g. former findings on lensed quenching galaxies by \citealt{2013ApJ...777...87G}).}. 
{Only space {or adaptive-optics} imaging will provide accurate constraints for these parameters and} confirm {these findings}, but meanwhile, to {better} assess the reliability of {the ground-based inferences}, we {test the procedure using   
20 mock ECs}. We follow the same approach used to simulate the lensing systems centered on randomly selected red luminous galaxies used to train our CNN on KiDS ground-based images (see Li+20). In particular we produce 
quad configurations using a source effective radius varying between $R_{\rm e}=$[0.05$''$, 0.15$''$],  $n-$index=[0.05,2], lens effective radius $R_{\rm e}=$[1.0$''$, 1.5$''$] and Einstein radius, $R_{\rm E}$=[1.0$''$, 1.5$''$]. {The ranges adopted are meant to cover the parameter space embracing the two crosses, and to demonstrate that the lens model tool can recover the parameters correctly. In particular we have tested the case of very small $R_{\rm e}$ and $n-$index of the source by modeling 4/20 mock lenses with $R_{\rm e}<0.07''$ and $n-$index$<0.6$.}
The simulated lenses are then convolved with a typical $r-$band PSF of KiDS observations and noise is finally added to produce realistic KiDS-like r-band EC images. 
We then run 
the \textit{lfit$\_$gui} using the same configuration file on these mock ECs and derived the lensing parameters like for the real ECs. The derived source $R_{\rm e}$ and $n-$index fall 
on the one-to-one relation with the input ones. To quantify this we derive the following quantities: 
$\Delta R_{\rm e}/R_{\rm e}=(R_{\rm e, in}-R_{\rm e, out})/R_{\rm e, in}=-0.01\pm0.06$ and $\Delta n/n=(n_{\rm in}-n_{\rm out})/n_{\rm in}=-0.01\pm0.15$. {The scatter is even smaller ($\sim 0.03$ both in $R_{\rm e}$ and $n$) for the 4 most extreme cases}. Since they are both consistent with zero and 
the ($1\sigma$) scatter is consistent with typical errors from the best-fit parameters in Table \ref{tab:lens_mod} ($\lsim 20\%$), we are confident 
that the ``compactness'' of the sources as well as its ``disk-like'' nature are real, although for the latter there might be more freedom about the exact value of the $n-$index. We also remark 
that sub-pixel sizes of the sources are refereed to the source plane, where \textit{lfit$\_$gui} maps the source model with a spatial resolution 10$\times$ higher than the pixel scale of the lens plane, where the source images are observed ($0.2''$ for KiDS images). Hence sizes smaller than a single pixel are well within the reach of the tool we use. {However, we have found that 0.04$''$ is the lowest limit for our groud-based observations. In fact, 
mock lenses having sources with $R_{\rm e}$ in the range [0.01,0.04] are recovered with too large uncertainties.
}


As {described }
in \S\ref{sec:MUSE}, the source spectra are dominated by absorption lines, which are typical of a relatively evolved population, and show only weak emission lines  
(see Fig. \ref{fig:spectra}), somehow at odds with the blue colors produced by their continuum. {In particular, KIDS-EC1 shows a low-SNR [OII] emission, while KIDS-EC2 shows 
[OII] and H$\epsilon$,  H$\delta$ and H$\gamma$  Balmer lines in emission, although the latter are superimposed to absorption Balmer lines, 
which makes any modeling of their profile very degenerate (e.g. continuum level, velocity dispersion of the absorption and emission lines, age, metallicity). Using the [OII] line, after some careful continuum subtraction, we estimate a tentative 
SFR of $\sim0.1 M_\sun/$yr (using Eq. 3 in \citealt{1998ARA&A..36..189K}). We stress that a SNR$\sim6$ is not high enough for a robust estimate. }

\begin{figure}
    \centering
    \includegraphics[width=10cm]{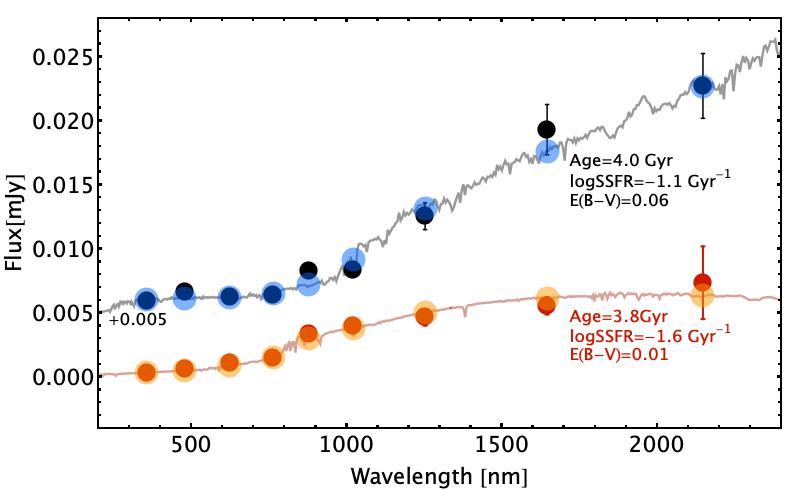}\vspace{3pt}
    \caption{SED fitting of the mean optical+NIR photometry obtained from averaging the de-lensed fluxes of the most magnified images in Table \ref{tab:EC_phot}. KIDS J232940-34092 photometry (black points with error bars) is plotted together with the best fit template model (gray solid line), corresponding to the parameters reported in the figure (see text for more details) {and solar metallicity}. Dark blue points show the corresponding photometry from the model used to fit the observations. KIDS J122456+005048 photometry (red points with errorbars), is plotted against the best fit model (light red line) and integrated photometry (yellow points), as well as model parameters {for the sub-solar metallicity case (see text for details)}. Data from KIDS J232940-34092 have been shifted +0.005 upward for clarity.}
    \label{fig:SED_sources}
\end{figure}

To {better} characterise \chiara{the nature of the two sources, we thus exploit their 9-band photometry (Table \ref{tab:EC_phot}) and run \cigale, as already done for the deflectors. }
In order to increase the 
\chiara{SNR} of the source SEDs, we 
average the fluxes of the higher magnified lensed images (AB for KIDS-EC1 and AC for KIDS-EC2) and obtain the de-lensed (i.e. using the $r-$band magnification, $\mu_{\rm r}$, as in Table~\ref{tab:lens_mod}) SED in Fig.~\ref{fig:SED_sources}. {We try 
two extreme metallicity scenarios: a standard solar metallicity and a largely sub-solar one ($Z=0.0004$), being the latter suggested by the strength of the Balmer lines compared with the Calcium H and K of KIDS-EC2 (for EC1 these lines are too redshifted to fall in the MUSE wavelength range). 
For KIDS-EC1 we obtain 
a better fit with solar metallicity (reduced $\chi^2=1.7$, corresponding to a $\sim$10\% significance for 8 degree-of-freedom), while for EC2 the fit is better 
for the sub-solar metallicity ($\chi^2=0.7$, i.e. $\sim$ 70\% significance). 
The best fits, shown in Fig.~\ref{fig:SED_sources}, give for 
KIDS-EC1  
an age of $4.0\pm0.1$ Gyr and a SFR of $9.1\pm1.2$ $M_\sun/$yr, while for for KIDS-EC2,  
age=$3.8\pm0.4$ Gyr and SFR$=0.4\pm0.1$ $M_\sun/$yr}\footnote{For completeness, the results obtained for KIDS-EC1 assuming a sub-solar metallicity are age=$4.0\pm1.0$ Gyr and SFR$=0.4\pm0.1$ $M_\sun/$yr and the results for KIDS-EC2 with solar metallicity are age=$1.2\pm0.2$ Gyr and SFR$=5.5\pm0.3$ $M_\sun/$yr. Note that full spectro-photometric  stellar population analysis 
of the sources is beyond the purpose of this letter and will be addressed in a separate paper.}. Hence both systems have an old age for their redshift, 
but are still forming stars at a low rate,  although we see only few emission lines in their spectra. In both cases the sources are quite massive: KIDS-EC1 has a stellar mass of $\log M_*/M_\sun=11.08^{+0.04}_{-0.05}$
and KIDS-EC2 of $\log M_*/M_\sun=10.21^{+0.02}_{-0.02}$
, hence their specific star formation rate ({\rm sSFR}) are $\log {\rm sSFR}/$Gyr$=-1.12\pm0.07$  and $\log {\rm sSFR}/$Gyr$=-1.61\pm0.03$ respectively, i.e. generally lower than the the typical values expected for the main sequence (MS) of star-forming (SF) galaxies at $z>1$ in the same mass range ($\log {\rm sSFR}_{\rm MS}/$Gyr$\sim0.0\pm0.4$, see e.g. \citealt{2015MNRAS.453.2540J}). 

Also for the lensed images we have performed a double check with \lephare\ and we have found  slightly lower masses (0.1 and 0.2 dex for KIDS-EC1 and EC2, respectively) but consistent or even lower $\log {\rm sSFR}/$Gyr ($-1.3$ and $-1$ respectively), hence confirming the quenching status of the two systems.

{\it Are these systems special?} They are for their sizes, as they are both outliers of the typical size--mass relation of SF systems by $\sim0.5$dex in effective radius. \citet{2017ApJ...834L..11A}, for instance, found a mean $\log R_{\rm e}/$kpc$\sim0.6$ and $0.5$ in the redshift bin $1<z<1.5$ for $\log M_*/M_\sun\sim11.1$ and $\sim10.2$ respectively, while we have $\log R_{\rm e}/$kpc$=-0.06$ for KIDS-EC1 and $-0.34$ for KIDS-EC2\chiara{. This means that they deviate}
significantly from normal galaxies at their redshift, while they are closer to typical sizes of SF galaxies at $z\gsim 5$. 
On the other hand, they also show very low {\rm sSFR},  deviating from the MS by $\Delta_{\rm MS}=\log$sSFR$_{\rm MS}-\log$sSFR$\sim-1.1$ and $-0.4$. Simulations from \citet{2016MNRAS.457.2790T} have shown that these low sSFR, together with sizes of $\log R_{\rm e}/$kpc$<0$, are typical of a ``post--blue nugget'' (BNs, hereafter) phase, i.e. systems having gone through compaction and entering their quenching phase. \citet{2018ApJ...858..114H} have also found that most of the massive, compact systems at z$>1$ ($\log M_*/M_\sun\gsim10.3$) tend to be in such a post--BN phase.   
  

\section{Conclusions and Perspectives} \label{sec:conclusions}
We have presented the \chiara{confirmation and modelling} of two Einstein Crosses \chiara{found in the KiDS footprint. }
The confirmation is based on MUSE spectroscopy, unequivocally showing the strong-lensing nature of the systems. We have detected the same spectral features in the four images of the two crosses and inferred a redshift of $z_{\rm s}\sim1.59$ for KIDS-EC1 and  $z_{\rm s}\sim1.10$ for KIDS-EC2, both higher that their respective deflector galaxies, i.e. two old early-type galaxies at $z_{\rm l}\sim 0.38$ (KIDS J232940-340922) and $z_{\rm l}\sim 0.24$ (KIDS J122456+005048).

The discovery is 
exceptional as \chiara{we only inspected $\sim1000$deg$^2$ so far and} Einstein Crosses (ECs) are very 
rare phenomena.  
However, general predictions on the number of expected quads (of which ECs are a special case) are based on the assumption that these are generated by quasars (see e.g. \citealt{2010MNRAS.405.2579O}). For the two ECs presented here, we have shown that they are produced by ultra--compact, blue, quenching galaxies. In particular:
\begin{enumerate}
\item to reproduce the cross configuration, the best lensing model (see Table \ref{tab:lens_mod}), assuming a singular isothermal ellipsoid (SIE) with external shear, 
predicts that the sources have very compact sizes: $R_{\rm e}\sim0.9$ kpc for KiDS-EC1 and $R_{\rm e}\sim0.5$ kpc for KiDS-EC2, i.e. $>10\sigma$ off the typical size-mass relation of normal SF galaxies at the same redshifts (see \citealt{2017ApJ...834L..11A});
\item  
the spectra of the sources show a dominance of absorption lines, typical of a quite evolved stellar population, which has been confirmed by SED fitting of the 9-band photometry 
\chiara{performed on the average of the two highest magnified images of each system. We inferred} 
old age\chiara{s} (4 Gyr and 3.8 Gyr for KIDS-EC1 and KIDS-EC2, respectively) and moderate star formation ($9.1$ and $0.4~M_\sun/$yr, respectively). 
However, 
the inferred stellar masses ($\log M_*/M_\sun\sim$11.08 and 10.21, respectively)
imply very low specific star-formation rates ($\log$sSFR$=-1.1$ and -1.6 Gyr$^{-1}$, respectively), typical of quenching galaxies at $z>1$ (see e.g. \citealt{2016MNRAS.457.2790T}, \citealt{2018ApJ...862..125N}).
\end{enumerate}
In \S\ref{sec:char_source}, we have argued that the combination of an extremely compact size and the low sSFR is compatible with the sources being two massive post--BNs. 
These are compact massive galaxies having almost exhausted their star-forming phase and currently undergoing quenching. As such, these systems are important to understand the transformation of primordial disks into the compact cores (``red nuggets'', see \citealt{2014MNRAS.438.1870D}) of today's large elliptical galaxies in the first phase of their evolution, before they enter their subsequent merging phase (e.g. \citealt{2012ApJ...744...63O}). 
These systems have been predicted to be very numerous in simulations (e.g. \citealt{2015MNRAS.450.2327Z}, \citealt{2016MNRAS.457.2790T}) and their census at $z>1$ has just started, including the confirmation of their abundance and physical properties (see e.g. \citealt{2018ApJ...858..114H}). Having observed two of such systems in a peculiar lensing configuration, might suggest 
that, indeed, they are not uncommon at $z>1$.
To make a rough estimate of the expected numbers of such EC events from compact post--BN systems, we can use predictions based on a size-luminosity relation compatible with high$-z$ studies calibrated over compact star-forming galaxies from \citet{2015ApJ...811...20C}, 
reproducing  typical ground based surveys with seeing $\leq 1''$\footnote{We have used the pre-compiled predictions for the Dark Energy Survey (DES), with seeing~1$''$ in github.com/tcollett/LensPop. These simulations are consistent with KiDS in terms of depth ($r\sim$25, 5$\sigma$ within 2$''$), {hence giving access to a similar lens luminosity distribution}, and only slightly worse in image quality ($\sim1''$ for DES vs 0.7$''$ for KiDS and a pixel size ~0.26$''$ vs 0.21$''$). Furthermore the adoption of a slightly larger seeing allows us to better account for the selection function introduced by the visual inspection, as the human eye tends to give a lower grade to arcs or multiple images that are too close to the lens center as they look diluted/confused in the lens starlight.}. We 
estimate $\sim1$ EC over 1000deg$^2$ (assuming $<0.1''$ alignment between source and deflector), generated by a slightly flattened mass distribution like the one estimated for our ECs ($q=$[0.7, 0.9]) by a compact source (size $\lsim 1$ kpc) at $z<1.7$, with sufficient SNR ($>10$) to be identified as a sure lens around bright deflectors ($r<21$, according to our selection in Li+20). 
This rough prediction can be likely an upper limit because we are assuming that all sources aligned within 0.1$''$ produce an EC, but using mock lenses we have checked that most of them should look as such (see \S\ref{sec:char_source}). The two ECs reported here are possibly slightly overabundant with respect to the expectation for standard star-forming systems. This is a first interesting indication that this population of post--BNs might be particularly abundant at $z>1$. {According to the same predictions, releasing any limitation on the brightness of the lens {(i.e. $r<25$)}, the expected EC/quad configurations are $\sim$ half a dozen for sources at $z<4$  (see also \S\ref{sec:intro})}. Hence we expect to confirm other systems in KiDS, in the future. 
More importantly we can expect that
for next generation surveys like LSST, EUCLID or CSST, 
we can discover $4000$ to $7000$ similar quad-like configurations. This will represent a unique opportunity to perform systematic studies of this population of compact systems in great details using lensing as a ``gravitational telescope'', which will be prohibitive to observe without lensing magnification (see e.g. \citealt{2017Natur.546..510T}) before extremely large telescopes will be online. 

\section*{Acknowledgements}
We thank the anonymous referee for the constructive reports that helped us to strengthen the results of our paper. NRN acknowledges financial support from the “One hundred top talent program of Sun Yat-sen University” grant N. 71000-18841229. RL acknowledge support from China Postdoctoral Science Foundation 2020M672935 and Guangdong Basic and Applied Basic Research Foundation 2019A1515110286. RL also acknowledge the Fundamental Research Funds for the Central Universities, Sun Yat-sen University 71000-31610034. CS is supported by a Hintze Fellowship at the Oxford Centre for Astrophysical Surveys. CT acknowledges funding from the INAF PRIN-SKA 2017 program 1.05.01.88.04. GD acknowledges support from CONICYT project Basal AFB-170002. KK acknowledges support by the Alexander von Humboldt Foundation. CH acknowledges support from the European Research Council under grant number 647112, and support from the Max Planck Society and the Alexander von Humboldt Foundation. HH is supported by a Heisenberg grant of the Deutsche Forschungsgemeinschaft (Hi 1495/5-1) and ERC Consolidator Grant (No. 770935). MB is supported by the Polish Ministry of Science and Higher Education through grant DIR/WK/2018/12, and by the Polish National Science Center through grants no. 2018/30/E/ST9/00698 and 2018/31/G/ST9/03388. We thank Yiping Shu for providing the lens fitting code \textit{lfit$\_$gui}, and for technical support. We also thank Felipe Barrientos and Nicolas Tejos for useful discussions. 

\appendix

\section*{Impact of the SIE assumption and the fitting tool}\label{sec:app1}
{
The mismatch between the stellar velocity dispersion measured from the spectrum and the lensing dispersion of KIDS-EC1 found in \S\ref{sec:lens_model} suggests that the assumption of a SIE mass density might be inappropriate. 
To investigate the impact of this assumption and check the reliability of the main parameters derived by \textit{lfit\_gui} and reported in Table \ref{tab:lens_mod}, we use a different lens model tool, \textit{LENSED} (\citealt{2016MNRAS.463.3115T}). 
We stress that a full comparison of different fitting techniques is beyond the purpose of this paper, however we can use the relevant outputs of an independent tool to validate the main results of this paper. 
First, as \textit{LENSED} allows a free slope for the mass density profile (i.e. an Elliptic Power Law, EPL, model), we can check how this deviates from SIE (3D slope$=2$). For KIDS-EC1 we find the best-fit 3D slope$=-1.75\pm0.06$, which is consistent with the \citet{2010ApJ...724..511A} formula ($\gamma'-2=2.67(f_{\rm SIE}-1)+0.20$, where $\gamma'$ is the 3D slope and $f_{\rm SIE}=\sigma_{\rm e/2}/\sigma_{\rm SIE}$) for the mismatch between stellar velocity dispersion measured from the spectrum and the lensing dispersion for the SIE model. For EC2 we have a 3D slope$=2.04\pm0.01$, which is instead very close to the 3D slope predicted for a SIE. 
Due to the different slopes, we derive different 2D mass extrapolation at $R_{\rm e}/2$: for KIDS-EC1 we have $\log M(R_{\rm e}/2)/M_\sun=10.99\pm0.07$, for KIDS-EC2 we have $\log M(R_{\rm e}/2)/M_\sun=11.20\pm0.02$. The inferred dark matter fractions are $f_{\rm DM}=0.54\pm0.11$ for KIDS-EC1 and $f_{\rm DM}=0.58\pm0.05$ for KIDS-EC2, i.e. fully consistent with the SIE results within $1\sigma$. 
This shows that the inner slope does not impact significantly our inference on the DM content of the galaxies, as well as the overall conclusions related to the lensing parameters in Table \ref{tab:lens_mod}, hence the use of the SIE in the rest of the paper is fairly justified.}

{Another central result of the paper is the compactness of the source, which, as discussed in \S\ref{sec:char_source}, is challenging to assess from ground based imaging and might depend, e.g., on the way a given tool samples the model at the sub-pixel scale, performs the convolution with the PSF and re-bins the model. We have double checked the results of \textit{lfit\_gui} against \textit{LENSED} and confirmed both compactness and the pseudo-exponential light profiles of the sources, even though the  central values are consistent only within $\sim 2\sigma$. In fact, from \textit{LENSED} we found that the source sizes of KIDS-EC1 is $R_{\rm e}=0.05''\pm0.01''$ while its $n-$index is $n=1.2\pm0.3$; for KIDS-EC2, we found instead $R_{\rm e}=0.03''\pm0.01''$ and $n=1.58\pm0.07$. {Finally, source axis ratios and magnitudes are fairly similar to \textit{lfit\_gui} inferences: for KiDS-EC1, \textit{LENSED} gives ($b/a,~r$) $=$(0.24, 23.54), for KiDS-EC2 (0.64, 23.80). As comparison, \textit{lfit\_gui} gives (0.5, 23.78), (0.7, 23.69) for KiDS-EC1 and KiDS-EC2, respectively.}
}



\begin{deluxetable*}{rrrlllllllll}
\tablecaption{Einstein Cross Optical-NIR photometry \label{tab:EC_phot}}
\tablewidth{0pt}
\tablehead{
\colhead{ID} & \colhead{$\Delta$RA} & \colhead{$\Delta$DEC} & \colhead{$u$} &\colhead{$g$}  & \colhead{$r$} & \colhead{$i$} & \colhead{$Z$} & \colhead{$Y$} & \colhead{$J$} & \colhead{$H$} & \colhead{$K_{\rm s}$} \\
\colhead{ } & \colhead{(arcsec)} & \colhead{(arcsec)} & \colhead{(mag)} & \colhead{(mag)}& \colhead{(mag)}& \colhead{(mag)}& \colhead{(mag)}& \colhead{(mag)}& \colhead{(mag)}& \colhead{(mag)}& \colhead{(mag)}
}
\footnotesize
\startdata
\multicolumn{12}{c}{\textsc{KIDS-EC1}: KIDS J232940-340922 RA$=352.417753$ DEC$=-34.156375$}\\ 
\hline
G & $0.00\pm0.03$ & $0.00\pm0.03$ &21.44&20.69&19.77&19.29&19.05&18.48&17.90&17.19&16.48\\
 &  &  &(0.22)&(0.09)&(0.04)&(0.04)&(0.12)&(0.06)&(0.08)&(0.08)&(0.04)\\
A & $-1.04\pm0.06$ & $-0.75\pm0.05$ & 22.57 & 21.97 & 22.29 & 22.27 & 21.16 & 21.14 & 20.46 & 19.70 & 19.39\\
 &  &  &(0.17)&(0.15)&(0.11)&(0.17)&(0.11)&(0.14)&(0.12)&(0.12)&(0.12)\\
B & $0.87\pm0.08$ & $0.99\pm0.05$  &22.65&21.98&22.36&22.10&21.32&21.29&20.24&19.60&19.44\\
 &  &  &(0.19)&(0.11)&(0.12)&(0.15)&(0.13)&(0.17)&(0.10)&(0.10)&(0.12)\\
C & $-0.58\pm0.09$ & $0.67\pm0.07$ &24.06&22.24&22.62&22.23&21.31&21.18&20.38&20.14&20.32\\
 &  &  &(0.79)&(0.17)&(0.29)&(0.37)&(0.24)&(0.21)&(0.16)&(0.23)&(0.39)\\
D & $0.61\pm0.19$& $-0.72\pm0.16$ &25.44&22.80&23.14&23.24&21.86&21.81&21.32&20.23&19.57\\
 &  &  &(1.65)&(0.39)&(0.38)&(0.51)&(0.39)&(0.71)&(0.42)&(0.43)&(0.18)\\
\hline
\multicolumn{12}{c}{\textsc{KIDS-EC2}: KIDS J122456+005048 RA$=186.233401$ DEC$=+0.846682$ }\\ 
\hline
G & $0.00\pm0.01$ & $0.00\pm0.01$ &20.94&19.30&18.10&17.48&17.20&17.04&16.67&16.24&15.98\\
 &  &  &(0.09)&(0.03)&(0.03)&(0.04)&(0.03)&(0.05)&(0.06)&(0.02)&(0.38)\\
A & $-1.24\pm0.05$ & $0.78\pm0.05$ &23.13&22.52&21.90&21.57&20.74&20.51&20.34&20.33&19.82\\
 &  &  &(0.14)&(0.16)&(0.17)&(0.13)&(0.08)&(0.09)&(0.11)&(0.08)&(0.42)\\
B & $0.50\pm0.07$ & $1.29\pm0.08$ &23.25&23.14&22.01&21.79&21.05&20.97&20.66&20.56&19.87\\
 &  &  &(0.11)&(0.33)&(0.21)&(0.21)&(0.09)&(0.11)&(0.12)&(0.09)&(0.42)\\
C & $1.27\pm0.06$ & $-0.96\pm0.06$ &23.44&22.63&21.97&21.65&20.73&20.62&20.43&20.08&19.96\\
 &  &  &(0.15)&(0.16)&(0.15)&(0.13)&(0.08)&(0.11)&(0.11)&(0.08)&(0.44)\\
D & $-0.73\pm0.09$& $-1.14\pm0.09$ &23.36&22.77&22.11&21.93&21.22&21.12&20.94&20.69&20.34\\
 &  &  &(0.18)&(0.29)&(0.22)&(0.27)&(0.15)&(0.16)&(0.19)&(0.10)&(0.46)\\
\enddata
\tablecomments{Objects coordinates are in degree, errors on magnitudes in different bands are given in brackets. Magnitudes of the lensed images (A, B, C, D) are total magnitudes obtained by Gaussian fit after the central galaxy G has been removed. Galaxy magnitude are total S\'ersic magnitudes.}
\end{deluxetable*}

\begin{deluxetable}{lrr}
\tablecaption{Summary of the Einstein cross main parameters \label{tab:lens_mod}}
\tablewidth{0pt}
\tablehead{
Parameter & KIDS-EC1 & KIDS-EC2
}
\startdata
\hline
\multicolumn{3}{c}{MUSE spectroscopy}\\ 
\hline
$z_{\rm l}$ & $0.3810\pm0.001$& $ 0.2372\pm0.0005$\\
$z_{\rm s}$ & $1.59\pm0.01$  & $1.10\pm0.01$\\
$\sigma_{\rm R_{\rm e}/2}$ (km~s$^{-1}$) & $192\pm4$ & $248\pm5$ \\
\hline
\multicolumn{3}{c}{Lensing model}\\ 
\hline
$\theta_{\rm E}$ (arcsec)& $0.99\pm0.02$ & $1.42\pm0.01$\\
$R_{\rm E}$ (kpc)& $5.2\pm0.1$ & $5.4\pm0.1$\\
$\sigma_{\rm E}$ (km~s$^{-1}$) & $226\pm8$ & $260\pm5$\\
lens $b/a$ & $0.91\pm0.07$ & $0.68\pm0.07$ \\
lens $PA$ & $1\pm56$ & $148\pm2$ \\
$\mu_r$ (A, B, C, D) & (3.6, 3.4, 2.7, 1.7) & (5.6, 5.1, 5.3, 4.6) \\
lens $PA$ & $1\pm56$ & $148\pm2$ \\
star $n-$index & $2.8\pm0.3$ & $3.43\pm0.08$\\
star $b/a$  & $0.89\pm 0.03$ & $0.59 \pm 0.01$\\
star $PA$ & $107\pm9$& $146\pm1$ \\
star $R_{\rm e}$ (arcsec) & $1.12\pm 0.10$ & $1.63 \pm 0.05$ \\
star $R_{\rm e}$ (kpc) & $5.8\pm 0.5$ & $6.2 \pm 0.2$ \\
$R_{\rm E}/R_{\rm e}$ & $0.90\pm 0.08$ & $0.87 \pm 0.03$ \\
star $\Delta$RA& $-0.02\pm0.02$ & $0.01\pm0.01$\\
star $\Delta$DEC& $0.02\pm0.02$ & $0.01\pm0.01$\\
source $n-$index & $0.4\pm0.1$ & $0.2 \pm 0.1$\\
source  $b/a$  & $0.5\pm 0.1$ & $0.7 \pm 0.1$\\
source  $R_{\rm e}$ (arcsec) & $0.10\pm 0.02$ & $0.06\pm 0.01$\\
source  $R_{\rm e}$ (kpc) & $0.87\pm 0.14$ & $0.46\pm 0.04$\\
source $\Delta$RA& $-0.03\pm0.02$ & $0.04\pm0.01$\\
source $\Delta$DEC& $-0.07\pm0.02$ & $0.05\pm0.01$\\
external shear  $\gamma_{\rm ex}$  & $0.25\pm 0.03$ & $0.01\pm 0.01$ \\
\hline
\multicolumn{3}{c}{Mass estimates}\\ 
\hline
 $\log M(R_{\rm E})/M_\sun$ & $11.28\pm0.02$ &  $11.42\pm0.01$ \\
$\log M(R_{\rm e}/2)/M_\sun$ & $11.04\pm0.03$ &  $11.18\pm0.01$ \\
$\log M_{\rm J}(R_{\rm e}/2)/M_\sun$ & $10.95\pm0.05$ & $11.21\pm0.02$\\
$\log M_{*}/M_\sun$ & $11.18\pm0.04$ & $11.33\pm0.03$\\
$\log M_{*}/M_\sun(R_{\rm e}/2)$ & $10.65\pm0.04$ & $10.81\pm0.03$ \\
$f_{\rm DM}(R_{\rm e}/2)$ & $0.60\pm0.07$  & $0.56\pm0.04$ \\
$\Delta f_{\rm DM}=f_{\rm DM,len}-f_{\rm DM,J}$ & $0.09\pm0.10$  & $-0.02\pm0.07$\\
\enddata

\tablecomments{Einstein cross parameters. Muse spectroscopy: redshift of the lens and source and velocity dispersion of the lens calculated at $R_{\rm e}/2$. Lensing models: Einstein radius in arcsec ($\theta_{\rm E}$) and kpc ($R_{\rm E}$) and the model SIE velocity dispersion, $\sigma_{\rm E}$, followed by self-explaining parameters related to the total mass (labelled by `lens'), stellar mass parameters (labelled by `star') and source parameters (labelled by `source') light distribution. Offsets ($\Delta$RA, $\Delta$DEC) are calculated with respect to the lens center. Mass estimates: summary of the mass estimates from lensing model ($M$), Jeans model ($M_J$), and stellar population $M_*$ (see text for details) together with the projected DM fractions ($f_{\rm DM}$) and difference between lens ($f_{\rm DM,len}$) and Jeans ($f_{\rm DM,J}$) analyses.}
\end{deluxetable}

\end{document}